\documentclass[
reprint,
amssymb,
 aps,
pre,
]{revtex4-2}
\usepackage{amsmath}
\usepackage{amsfonts}
\usepackage{color}
\usepackage{xcolor}
\usepackage{graphicx}% Include figure files
\usepackage{dcolumn}% Align table columns on decimal point
\usepackage{natbib}
\usepackage{color,soul}
\setstcolor{red}
\sethlcolor{yellow}
\usepackage{comment}
\usepackage{tikz}
\usetikzlibrary{shapes}
\usepackage{url}
\usepackage[labelformat=simple]{subcaption}

\usepackage{graphicx}
\newcommand{\markerwisteria}{\raisebox{0.5pt}{\tikz{\node[draw,scale=0.8,regular polygon, regular polygon sides=4,fill=blue!35!white](){};}}}
\newcommand{\markerpurple}{\raisebox{0.5pt}{\tikz{\node[draw,scale=0.8,regular polygon, regular polygon sides=4,fill=white!30!pink](){};}}}
\newcommand{\markerviolet}{\raisebox{0.5pt}{\tikz{\node[draw,scale=0.8,regular polygon, regular polygon sides=4,fill=violet](){};}}}
\newcommand{\markergray}{\raisebox{0.5pt}{\tikz{\node[draw,scale=0.8,regular polygon, regular polygon sides=4,fill=gray](){};}}}
\newcommand{\markerbeige}{\raisebox{0.5pt}{\tikz{\node[draw,scale=0.8,regular polygon, regular polygon sides=4,fill=brown!20!white](){};}}}

\begin{document}
\title{Kullback-Leibler cluster entropy to quantify volatility correlation and risk diversity}
\author{L. Ponta}
\affiliation{Università di Genova, Italy}
\author{A. Carbone}
\affiliation{Politecnico di Torino,  Italy}
%\pacs{05.40.-a}{Fluctuation phenomena, random processes, noise, and Brownian motion}
%\pacs{89.70.Cf}{Entropy and other measures of information}
%\pacs{89.65.Gh}{Economics; econophysics, financial markets, business and management}
%
\begin{abstract}
The  \textit{Kullback-Leibler cluster entropy}  $\mathcal{D_{C}}[P \| Q] $   is evaluated for the empirical  and model probability  distributions $P$  and $Q$ of the clusters formed in the  \textit{realized volatility} time series of five assets (SP\&500, NASDAQ, DJIA, DAX, FTSEMIB).   
The Kullback-Leibler functional  $\mathcal{D_{C}}[P \| Q] $  provides complementary perspectives about the stochastic volatility process compared to the Shannon functional $\mathcal{S_{C}}[P]$. While $\mathcal{D_{C}}[P \| Q] $ is maximum at the short time scales,  $\mathcal{S_{C}}[P]$ is maximum at the large time scales leading to complementary optimization criteria tracing back respectively  to the maximum   and  minimum relative entropy evolution principles. The realized volatility is modelled as a time-dependent fractional stochastic process characterized by power-law decaying  distributions with positive correlation ($H>1/2$).  As a case study, a multiperiod portfolio built on diversity indexes derived from the Kullback-Leibler entropy measure of the \textit{realized volatility}. The portfolio is robust and exhibits better performances over the horizon periods. A comparison with the portfolio built either according  to the uniform distribution or in the framework of the Markowitz theory is also reported.
\end{abstract}
\maketitle
\section{Introduction}
\label{sec:Introduction}
\textit{Information theoretic concepts}  based on the Kullback-Leibler functionals are  increasingly finding applications in several contexts \cite{vedral2002role,anderson2004model,roldan2012entropy,ichiki2023performance,gambhir2022learning}.  
Insightful perspectives on the dynamics underlying  stochastic processes can be taken in terms of  entropy evolution   \cite{mackey1989dynamic}. According to the second law of thermodynamics, the Shannon-Gibbs entropy $S$  increases over time and reaches a \textit{maximum value} $S_{max}$  corresponding to a steady-state. %or more (metastable) states
Entropy functionals are  expressed in terms of  probability distribution functions $P$ of the thermodynamic state. However, probability distributions are barely defined when empirical real world data are concerned.  In such circumstances, the procedure of \textit{coarse-graining} is implemented by partitioning the phase space into discrete not overlapping cells $C_j$.
%%%% $\bigcup_j C_j=X \quad$ and not overlapping with each others $\quad C_j\bigcup C_j'=\varnothing \, \text{for} \, {j \neq j'}$. 
Measures $\mu$ can be defined over the cell $j$ satisfying the condition  $0<\mu\left(C_j\right)<\mu(X)$ for all $j$. The \textit{coarse-grained distribution}  $P(\mu (C_j))$   is entered in the entropy functional yielding a so called \textit{coarse-grained entropy} $\cal S$ equal or larger than $S$ the entropy analytically defined  for the same system. 
\par
Time series  of random variables $x_t$  provide relevant observables of the system dynamics at subsequent  time instances $t$.  Coarse-graining  is obtained by cutting sequential segments  $x_{t^{\prime}}=x_t x_{t+1} \ldots x_{t^{\prime}-1}, t<t^{\prime}$ of varying lengths out of the time series $x_t$  \cite{crutchfield2012between,grassberger1983characterization}.  Then the entropy evolution  for the coarse grained  sequence $x_t$ is estimated  at varying \textit{block size}.  Such coarse-grained information measure is commonly referred to as \textit{block entropy}.   Coarse-graining  via  \textit{density-based clustering} consists in intersecting empirical  and thresholding data sets. The intersections define the clusters (cells) which are not affected by the drawbacks of the  standard \textit{center-based} clusters.  Sample by sample estimates of the probability distribution of  \textit{density-based cluster features}   are increasingly proved useful  when large amount of data should be investigated.     
Entropy measures implemented on {density-based cluster probabilities} have emerged as a promising area of complexity science and statistical machine learning (\textit{Information theoretic clustering}) \cite{gokcay2002information,jain2010data,vinh2010information,rodriguez2014clustering,BAILEY2023102245}. 
\par
The  \textit{cluster entropy} approach adopted in this work operates via  a density-based partition obtained by intersecting empirical data (e.g. a time series) with a pointwise threshold defined by a time dependent average \cite{carbone2004analysis,carbone2007scaling}. The extent  to which the cluster entropy exceeds the value expected by the analytic Shannon functional has been discussed in \cite{Carbone2013information} where the approach has been implemented on  human chromosome sequences. The  \textit{Kullback-Leibler cluster entropy} \cite{carbone2022relative} estimates the divergence  between the  probability distributions $P$ and $Q$ respectively of  the empirical and artificial cluster sequences taken as a model. 
How the \textit{Kullback-Leibler cluster entropy} can be used to infer the  optimal  distribution $P$  has been demonstrated for fractional Brownian processes (fBm) with $H$ the Hurst exponent    in the range  $ H \in [0,1]$  \cite{carbone2022relative}.  Keeping in mind the general property of the relative entropy  to reach its maximum value at short time scales and evolve towards  the minimum (zero) for $P=Q$ \cite{anderson2004model, bavaud2009information}, 
the Kullback-Leibler cluster entropy  complements the Shannon cluster  entropy,  a proxy of correlation  at large time scales. 
\par 
In  economics and finance, entropy tools are adopted for  asset pricing models, 
risk assessment and wealth allocation motivated by the concept that entropy itself is a measure of diversity  \cite{backus2014sources,ghosh2017what,ormos2014entropy,bera2008optimal,tumminello2007kullback}.  
Contrarily to the time series of prices  widely recognised to behave as simple Brownian motions with Hurst exponent  $H \sim 1/2$, the mechanisms of long-term correlation underlying the  stochastic dynamics of the volatility (i.e. the time series corresponding roughly to the local variance of the price time series)  are still debated.
Volatility  has been modelled  as a positively correlated fractional Brownian motions  with $H > 1/2$  \cite{comte1998long}. An anticorrelated stochastic process with 
 $H < 1/2$ has been proposed  with the  market prices being a semimartingale with $H \sim 1/2$ \cite{gatheral2018volatility}.
Numerical experiments reported in~\cite{cont2024rough} show that the  volatility exhibits  Hurst exponent $H < 1/2$ though its microscopic origin might be an artifact.  
%Research aimed to carefully examining the correlation of the stochastic volatility and  quantifying the associated Hurst exponent is ongoing.
This work intends to contribute to the debate by exploiting the inferential ability of the Kullback-Leibler cluster entropy.  To this scope the probability distributions estimated respectively in the \textit{realized volatility} series of five financial assets and in  fully uncorrelated Brownian motions are compared to each other.  
\par
\textit{Diversity measures} derived from various entropy functionals have been proposed   to quantify compositional heterogeneity of natural and man-made complex systems  \cite{hill1973diversity,leinster2021maximum,errunza2024learning}. The measures adopt the \textit{maximum entropy} as a proxy of  \textit{maximum diversity} consistently with the  spreading of the distribution  compared to the uniform or delta ones \cite{leinster2021maximum}. The concept of diversity has been originally adopted in finance in the portfolio theory framework pioneered by Markowitz   \cite{markowitz52portfolio,hakansson1971multi,steinbach2001markowitz}.
First attempts of using information theoretic tools for risk assessment and wealth allocation consisted in the introduction of the Markowitz mean-variance weights  $w_i$   into the Shannon entropy functional:
\begin{equation}
S(w_i)=-\sum_{i=1}^{{\cal{A}}} w_{i} \log w_{i} \quad ,
\label{eq:Shannonweights}
\end{equation}
where ${\cal{A}}$ is the number of assets. 
Eq.~(\ref{eq:Shannonweights}) ranges between the maximum value  $S(w_i) = \log {\cal{A}} $  when $w_{i}=1/{\cal{A}}$  (\textit{equally weighted assets}) and the minimum $S(w_{i})=0$  when $w_{i}=1$ for one asset and $w_{i}=0$ for all the others.
The Kullback-Leibler  entropy functional has been used to quantify the divergence of the Markowitz weights $w_{i}$ with respect to the uniformly distributed weights $u_{i}=1/{\cal{A}}$: 
 \begin{equation}
D(w_i,u_i)=\sum_{i=1}^{{\cal{A}}} w_{i} \log \frac{w_{i}}{u_{i}}\quad .
\label{eq:Kullbackweights}
\end{equation}
Though the  entropy functionals could mitigate strong fluctuations and  biases towards riskiest markets, the relationships~(\ref{eq:Shannonweights},\ref{eq:Kullbackweights})  still embed the  weights $w_{i}$ estimated  under the wrong belief of stationary normally distributed return.
Diversity measures suitable to capture the complexity of composite micro- and macro-economic features  still remains an open challenge  \cite{errunza2024learning}. 
The Kullback-Leibler cluster entropy will be used to define  a robust and sound set of diversity measures built upon the coarse grained probability  distribution of the  realized volatility of  tick-by-tick data of the  Standard \& Poor 500 (SP\&500), Dow Jones Ind. Avg (DJIA), Deutscher Aktienindex (DAX),  Nasdaq Composite (NASDAQ), Milano Indice di Borsa (FTSEMIB) assets.
\par
The manuscript is organized as follows.   Section \ref{subs:ClusterEntropy} includes the main notions and computational steps underlying the Kullback-Leibler cluster divergence approach.  Section  \ref{subs:portfolio} describes the construction of the diversity measures in terms of the relative cluster entropy. Section \ref{Sec:Results} illustrates the approach on volatility series of  tick-by-tick data of the  Standard \& Poor 500 (SP\&500), Dow Jones Ind. Avg (DJIA), Deutscher Aktienindex (DAX),  Nasdaq Composite (NASDAQ), Milano Indice di Borsa (FTSEMIB) assets.
The Kullback-Leibler cluster entropy and the diversity indexes are estimated  over twelve monthly periods covering the year 2018. A comparison with the measures obtained by using the Shannon cluster entropy,  the equally-weighted and the Sharpe ratio  estimates are also included. In Section \ref{Sec:Discussion}  the implications of the study are shortly discussed. A numerical example of investment is also provided. Conclusions are drawn in Section \ref{Sec:Conclusions}.
%%%%%%%%%%%%  METHODS   %%%%%%%%%%%%
\section{Methods}
\label{Sec:Methods}
A few definitions and the main computational steps of  the  \textit{Kullback-Leibler cluster entropy}  approach are summarized  in subsection \ref{subs:ClusterEntropy}.  The construction of the  \textit{Kullback-Leibler diversity index} is reported in    subsection \ref{subs:portfolio}.
\subsection{Kullback-Leibler Cluster Entropy}
\label{subs:ClusterEntropy}
As mentioned in the introduction, the evaluation of the entropy functional for real-world problems requires  the procedure of partitioning the data set (coarse-graining)  to yield the coarse-grained probability distribution of cells $C_j$. 
\par
In this work, the focus is on empirical time series of the realized volatility where the partition is obtained by intersecting the data with a  pointwise threshold \textit{density-based clustering}.
%%%%%%%%%%  Figure 1 %%%%%%%%%
\begin{figure*}%[]
\begin{subfigure}[]{0.329 \textwidth}
\includegraphics[width=\textwidth]{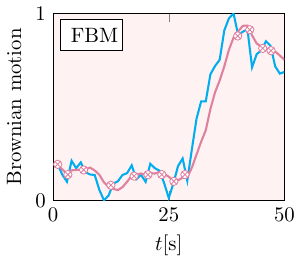}
\caption{}\label{subfig:a}
\end{subfigure}
\begin{subfigure}[]{0.329\linewidth}
\includegraphics[width=\textwidth]{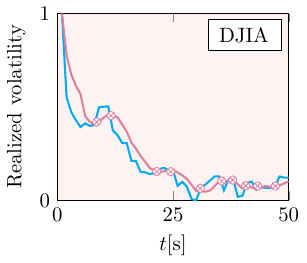}
\caption{}\label{subfig:b}
\end{subfigure}
\begin{subfigure}[]{0.329\linewidth}
\includegraphics[width=\textwidth]{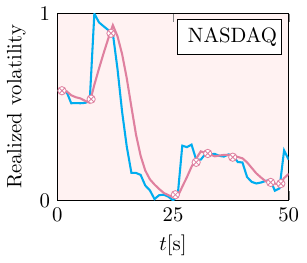}\\
\caption{}\label{subfig:c}
\end{subfigure}
\begin{subfigure}[]{0.329\linewidth}
\includegraphics[width= \textwidth]{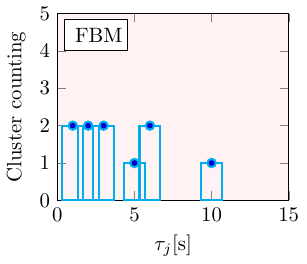}
\caption{}\label{subfig:d}
\end{subfigure}
\begin{subfigure}[]{0.329\linewidth}
\includegraphics[width= \textwidth]{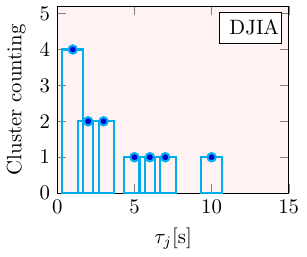}
\caption{}\label{subfig:e}
\end{subfigure}
\begin{subfigure}[]{0.329\linewidth}
\includegraphics[width=\textwidth]{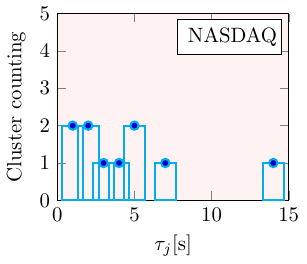}
\caption{}\label{subfig:f}
\end{subfigure}
\caption{\label{fig:kullbackshort} \textbf{Concept sketch of the Kullback-Leibler cluster method}. Top Panels: samples of the \textbf{(a)} model time series - an artificially generated Brownian motion with $H=0.5$ (FBM),  of the realized volatility series of \textbf{(b)} DJIA   and \textbf{(c)} NASDAQ assets (blue lines).  Sample lengths is $N=50$. The moving average with $n=5$ is plotted for each time series (purple lines).  The clusters, formed at the intersections of each series and moving average, are indicated by crossed circles. Bottom Panels: The clusters formed in (a),(b) and (c) are counted according to their length. The cluster occurrences are plotted in the bottom panels respectively for the FBM  \textbf{(d)},  DJIA \textbf{(e)}  and  NASDAQ \textbf{(f)}. The cluster counts represent the archetypes of the probability distribution function $P$ and $Q$ to be introduced in the Kullback-Leibler expression (Eq. (\ref{Kullbackdtau})).
}
 \end{figure*}
%%%%%
Consider a time series  $\{x_t \}$ of length $N$.  The local average \begin{equation}\widetilde{x}_{t,n} = \frac{1}{n} \sum_{n' = 0}^{n - 1} x(t-n') \end{equation} generates a family of time series of length $N-n$  for $n \in(1,N)$. It  is  $\widetilde{x}_{t,n}=\{x_t \}$  for $n=1$ and  $\widetilde{x}_{t,n}=\text{constant}$ for $n=N$. 
 Consequently, a family of partition  $\{ {\mathcal{C}}\} = \{ {\mathcal{C}_{n,1}},  {\mathcal{C}_{n,2}}, \ldots,  {\mathcal{C}_{n,j}},  \ldots\} $  of non-overlapping clusters is generated by the  intersections of $\{x_t \}$ and
$\{\widetilde{x}_{t,n}\} $. Each cluster  ${\mathcal{C}_{n,j}}$ is defined by the condition  $\epsilon_{t,n}={x_{t}}- \widetilde{x}_{t,n}=0$ for each $n$. The clusters  are characterized by the \textit{duration}
$\tau_j\equiv  \|t_{j}-t_{j-1}\|$,
with the instances $t_{j-1}$ and $t_j$ corresponding to subsequent  intersection pairs. 
The coarse grained distribution of the    cluster durations $P(\tau_j,n)$  is obtained by ranking the number of clusters ${\mathcal N}(\tau_j,n)$ according to their duration $\tau_j$:
\begin{equation}
P(\tau_j,n)=\frac{{\mathcal N}(\tau_j,n)}{{\mathcal N_C}(n)}
\end{equation}
with  ${\mathcal N_C}(n)=\sum_{j}^{}  {\mathcal N}(\tau_j,n)$  the  number of clusters generated  for each value of the parameter $n$.  The total number of clusters for all the possible values of the parameter $n$ is defined as:
${\mathcal N_C}=\sum_{n=1}^{N}{\mathcal N_C}(n)$. 
The normalization condition holds as usual:
\begin{equation}
\sum_{n=1}^N \sum_{j=1}^{{\mathcal N_C}(n)} P(\tau_j,n)= 1 \hspace{5pt}.
\end{equation}
\par
Next consider a coarse grained distribution function $Q(\tau_j,n)$ generated by a model of the time series $\{x_t \}$. The  \textit{relative cluster entropy} or \textit{cluster divergence} can be defined  in terms of the Kullback-Leibler functional of the  probabilities $P(\tau_j,n)$ and $Q(\tau_j,n)$:
\begin{equation}
\mathcal{D}_{j,n}[P || Q ] =    P(\tau_j, n)\log \frac{P(\tau_j, n)}{Q(\tau_j, n)} \hspace{7pt},
\label{Kullbackdtau}
\end{equation}
with the  condition $\mathrm{supp} (P) \subseteq \mathrm{supp}(Q)$.
By summing ~(\ref{Kullbackdtau})   over all the  cluster durations $\tau_j$ and all the partitions generated by varying the  parameter $n$, one has:
\begin{equation}
%\begin{align}
\mathcal{D_{C}}[P|| Q]   = \sum_{n=1}^N \sum_{j=1}^{{\mathcal N_C}(n)} P(\tau_j, n)\log \frac{P(\tau_j, n)}{Q(\tau_j, n)} \hspace{5pt}.
\label{Kullbackdtaun}
%\end{align}
\end{equation}
which quantifies the information yield when the empirical probability distribution $P$ is compared to the model probability distribution $Q$  \cite{carbone2022relative}.
\par
For illustrative purposes, short samples (50 points) of time series are plotted in Figs.~\ref{fig:kullbackshort}(a),(b),(c).  The time series taken as a model  is a Fractional Brownian motion (FBM) with Hurst exponent $H=0.5$ (blue line in Fig.~\ref{fig:kullbackshort} (a)). The empirical time series  to be analysed are the realized volatility of  DJIA  (blue line in Fig.~\ref{fig:kullbackshort} (b)) and NASDAQ (blue line in  Fig.~\ref{fig:kullbackshort} (c)). The intersection (crossed circles) with the moving average curves (purple lines) generate clusters of different lengths. The clusters are counted according to their lengths to  build the prototypical probability distributions to be entered in the Eqs.~(\ref{Kullbackdtau},\ref{Kullbackdtaun}) and are respectively shown in   Figs.~\ref{fig:kullbackshort}(d),(e),(f). 

%%%%%%%%  TABLE %%%%%%%
\begin{table*}%[h]
\footnotesize
\centering
\begin{tabular}{lccc|rrrrrrrrrrrr}
\toprule
{} & { (a)}  & { (b)} & { (c)} & {$p_1$} & {$p_2$}  & {$p_3$}  & {$p_4$} & {$p_5$}  & {$p_6$} & {$p_7$} & {$p_8$}  & {$p_9$}  & {$p_{10}$} & {$p_{11}$}  & {$p_{12}$}  \\
\hline
\hline
    NASDAQ & 2570		& 6982017  &	USD	    &	7007	&	7386	&	7181	&	6870	&	7131	&	7554	&	7568	&	7707	&	8091	&	8037	&	7434	&	7442	\\
S\&P500	& 505  & 6142443	 &	USD 	   &	2696	&	2822	&	2678	&	2582	&	2655	&	2735	&	2727	&	2813	&	2897	&	2925	&	2740	&	2790	\\

DJIA	& 30 	  & 5749145  &	USD	    &	24824	&	26187	&	24609	&	23644	&	24099	&	24635	&	24307	&	25334	&	25952	&	26651	&	25381	&	25826	\\
DAX	&30	& 7859601    &	USD   &	26	&	28	&	25	&	25	&	25	&	25	&	24	&	25	&	24	&	24	&	22	&	22	\\
FTSEMIB	& 40 	& 11088322 &	USD  &	28268	&	27576	&	27684	&	27697	&	28462	&	24920	&	25689	&	24668	&	23750	&	22349	&	21941	&	21460	\\
% FTSEMIB	&euro	  &	23430	&	22167	&	22746	&	22930	&	24159	&	21356	&	21926	&	21091	&	20448	&	19256	&	19258	&	18742	\\
\end{tabular}
\caption{\textbf{Asset Data}. Member firms constituting the asset (a).  Number of price ticks for the year 2018 (b). Currency (c).  Adjusted close price on the first available  day for each month of the year 2018 ($p_1 \ldots p_{12}$).}\label{tab:data}
\end{table*}
\par
\par
To further  illustrate how the relative cluster entropy operates,  Eq.~(\ref{Kullbackdtaun}) is written in terms of continuous  variables:
\begin{equation}
D_{C}[P|| Q] =\int P(\tau) \log \frac{P(\tau)}{Q(\tau)} d \tau \hspace{5pt}.
\label{Kullbackcint}
\end{equation}
For power-law probability distributions:
\begin{equation}
P(\tau)=\frac{\alpha_1-1}{\tau_{\min }}\left(\frac{\tau}{\tau_{\min }}\right)^{-\alpha_1}
\quad 
Q(\tau)=\frac{\alpha_2-1}{\tau_{\min }}\left(\frac{\tau}{\tau_{\min }}\right)^{-\alpha_2}
\end{equation}
%%%
where ${\alpha_1>1}$ and  ${\alpha_2>1}$ are the correlation exponents  and $\tau \in [1,\infty ]$.
After integration between  $\tau_{\min}=1$ and $\tau_{\max}=\infty$, Eq.~(\ref{Kullbackcint}) writes:
\begin{equation}
D_{C}[P|| Q]=\log \frac{\alpha_1-1}{\alpha_2-1} -\frac{\alpha_1-\alpha_2}{\alpha_1-1}
\label{DCI}
\end{equation}
where the condition $D_{C}(\infty)=0$ has been used to evaluate the integration constant.
The   cluster formation process (coarse-graining) based on the intersection of a time series $x_t$  and the moving average  $\widetilde{x}_{t,n}$ described at the beginning of this section  can be further understood in the general context of  \textit{first passage problems} \cite{redner2001guide} with  particular reference to the case of  fractional Brownian motions \cite{ding1995distribution}.
The probability distribution functions  of the cluster duration $\tau$  is a power-law with  exponent $\alpha=2-H$ where $H \in [0,1]$ is the Hurst exponent of the fractional Brownian motion generating the clusters \cite{carbone2004analysis}. Thus by using  $\alpha_1=2-H_1$ and  $\alpha_2=2-H_2$,  Eq.~(\ref{DCI}) writes:
\begin{equation}
{D_{C}}[P|| Q] =    \log \frac {1-H_1}{1-H_2}
+   \frac{ H_1 -H_2 }{1-H_1}\hspace{5pt}.
\label{Kullbackc00}
\end{equation}
%%%%%%
Eq.~(\ref{DCI})  turns out to depend only on the exponents  $\alpha_1$ and $\alpha_2$ of the power law distributions of the empirical  and model sequences.
Eq.~(\ref{Kullbackc00})  turns out to depend only on $H_1$ and $H_2$, respectively the Hurst exponent of the empirical and model sequences. 
\par
${D_{C}}[P|| Q]$  satisfies the general property of the relative entropy to be positive defined:
\begin{equation} {D_{C}}[P|| Q] \geq 0
\end{equation} over the whole range of  $\alpha_1=2-H_1$ and $\alpha_2=2-H_2$.    The minimum value:
\begin{equation}
{D_{C}}[P|| Q] = 0
\end{equation} is obtained for $H_1=H_2$ ($\alpha_1 =\alpha_2$).
\par
Eqs.~(\ref{DCI},\ref{Kullbackc00}) could be
% written as  \begin{equation} {D_{C}}[P|| Q]={D_{C}}[H_1|| H_2]\end{equation} and
interpreted  as an inferential measure of the correlation exponent $\alpha$ and of the Hurst exponent $H$ for fractional Brownian motions. 
The functionals $\mathcal{D}_{j,n}[P || Q ]$ defined in  Eq.~(\ref{Kullbackdtau}) are the individual components of the relative cluster entropy. $\mathcal{D}_{j,n}[P || Q ]$ are  dependent on the Hurst exponent  and can be separated into  positive or negative contributions respectively for positive  and  negative  correlated  sequences. It is worth remarking that the functionals $\mathcal{D}_{j,n}[P || Q ]$ depend on the cluster time scale $\tau_j$.

\begin{figure*}%[h]
\begin{subfigure}[]{0.329 \textwidth}
\includegraphics[width= \textwidth]{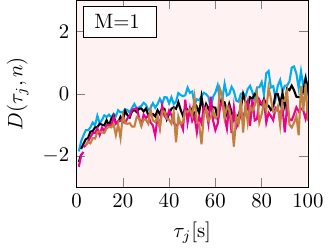}
\caption{}\label{subfig:a}
\end{subfigure}
\begin{subfigure}[]{0.329 \textwidth}
\includegraphics[width= \textwidth]{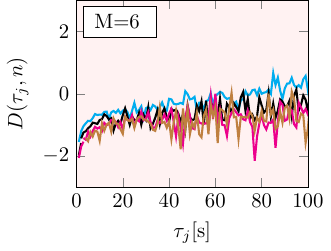}
\caption{}\label{subfig:b}
\end{subfigure}
\begin{subfigure}[]{0.329 \textwidth}
\includegraphics[width=\textwidth]{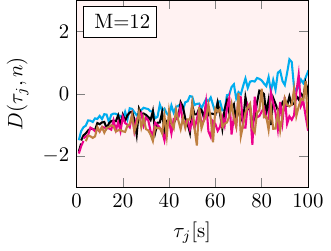}
\caption{}\label{subfig:c}
\end{subfigure}
\caption{\label{fig:kullbackentropy} \textbf{Kullback-Leibler cluster entropy functional $\mathcal{D}(\tau_j,n) $ vs cluster duration}.  Each curve  is estimated by means of  Eq.~(\ref{Kullbackdtau}) the Kullback-Leibler functional of ${{P}}(\tau_j, n)$  the  empirical probability distribution of the size of the clusters of the volatility series of the market S\&P500.   $Q(\tau_j,n)$ is the  probability distribution of the cluster size of  a synthetic fractional Brownian motion  with $H=0.5$.
Time series have length $N=492023$. Volatility is estimated according to Eq.~(\ref{eq:volatility}) with window $T = 180s$ for the three graphs. Plots refer respectively to  one (${\cal{M}}=1$){(a)}, six (${\cal{M}}=6$) {(b)}  and twelve  (${\cal{M}}=12$) {(c)} monthly periods sampled out of the year 2018. Different curves in the same plot refer to different values of  the moving average
window  $n$ ranging from  $50s$ to  $200s$ with step $50s$.}
\end{figure*}
%%%%%%%%%%%%%%%%%%%%%%
\subsection{Kullback-Leibler cluster diversity index}
\label{subs:portfolio}
%%%%%%%%%%%%%%%%%%%%%%
\par \medskip 
The \textit{relative cluster entropy index}  $I_{\cal {D}}$  can be defined in terms of the  Eqs.~(\ref{Kullbackdtau},\ref{Kullbackdtaun}) as: 
\begin{equation}
I_{\cal {D}} = \sum_{n=1}^N\sum_{j=1}^m D(\tau_j, n) + \sum_{n=1}^N \sum_{j=m}^N D(\tau_j, n) \quad.
\label{eq:Dindex} 
\end{equation}
The first term on the right hand side of Eq.~(\ref{eq:Dindex}) refers to the sum of  $D(\tau_j, n)$ over cluster lifetimes in the range  $1<\tau_j<\tau_m < n$ where  $D(\tau_j, n) \neq 0$ varies from the maximum for  $\tau_j \rightarrow 1$ to the minimum value $D(\tau_j, n) \rightarrow 0$ for  $\tau_j  = \tau_m  \approx n$.  The second term corresponds to  cluster lifetimes in the range  $n<\tau_j< N$  where  $D(\tau_j, n) \approx 0$. 
\par
$I_{\cal {D}}$  is a measure of diversity, suitable to  compare  cluster probabilities generated in different sequences. To this purpose, the index 
$I_{\cal {D}}$ must be normalized over the ensemble to be compared. The normalized indexes (\textit{Kullback-Leibler weights})  are defined as follows:   
\begin{equation}
\label{eq:Dweights}
{w}_{i,\cal {D}}=
\frac{I_{i,\cal {D}}^{-1}}
{\sum_{i=1}^{{\cal{A}}} 
I_{i,\cal {D}}^{-1}} \quad ,
\end{equation}
that satisfy  the conditions $\sum_{i=1}^{{\cal{A}}}{w}_{i,\mathcal{D}}=1$  and  $w_{i,\mathcal{D}} \geq 0$. 
The weights ${w}_{i,\mathcal{D}}$  allow to compare and rank time series with reference to an artificial time series taken as a model.  

\par \medskip
In section \ref{Sec:Results}, the relative cluster entropy approach will be used to quantify the divergence of the correlation exponents of the realized volatility series of five  assets.  Eqs.~(\ref{Kullbackdtau},\ref{Kullbackdtaun})  will be used to quantify  the correlation exponent $H_1$ of the empirical sequence   in comparison to a fractional Brownian motion with assigned Hurst exponent $H_2$ taken as a model.
 We will also  estimate  the \textit{relative cluster entropy index}  $I_{\cal {D}}$ and build a numerical example of a portfolio with the cluster weights ${w}_{i,\mathcal{D}}$  given in Eq.~(\ref{eq:Dweights}) to five financial assets.

%%%%%%%% 
%%%%%%%% RESULTS %%%%%%%%
%%%%%%%% 
\section{Results} 
\label{Sec:Results}
The \textit{relative cluster entropy} approach  will be implemented to analyse  the correlation degree of the realized volatility series of five assets. 
For  the sake of clarity, the definitions of return  and realized volatility  are shortly recalled. 
Given a time series of market price $p_t$, the \textit{return time series}   is defined as:
\begin{equation} \label{eq:return}
r_t = \log p_t - \log p_{t-1} \quad.
\end{equation}
The  \textit{realized volatility time series}  is defined as:
\begin{equation} \label{eq:volatility}
\sigma_{t,T} = \sqrt{\frac{\sum_{t=k}^{k+T}{(r_t-\mu_{t,T})^2}}{T- 1} } \hspace{15 pt},
\end{equation}
where $T$ is the volatility window and $\mu_{t,T}$ is the time series of the \textit{expected return} over $T$:
\begin{equation} \label{eq:returnm}
\mu_{t,T} = \frac{1}{T} \sum_{t=k}^{k+T} r_t \hspace{15 pt}.
\end{equation}
Eq.~(\ref{eq:volatility}) corresponds to the estimate of  the variance of  return  over the volatility window $T$  \cite{barndorff2002econometric,andersen2003modeling}.
\par
The analysis  are  carried on  tick-by-tick data of the  S\&P500, NASDAQ, DJIA, DAX and  FTSEMIB assets (details in Table \ref{tab:data}). Short samples of length $N=50$  of the realized volatility time series of the  DJIA and NASDAQ assets are plotted in Figs.~\ref{fig:kullbackshort}(b),(c).
The model series is a fractional Brownian motion FBM with $H=0.5$. A short sample of length $N=50$ of the FBM is plotted in Fig.~\ref{fig:kullbackshort}(a).
 Blue lines represent the time series $x_t$, purple line represents the moving average $\widetilde{x}_{t,n}$ with $n=5$.  The clusters are formed by the intersections of the series and its moving average. The  crossing points are drafted as \textit{otimes} symbols in Figs.~\ref{fig:kullbackshort} (a),(b),(c).   The clusters are counted and ranked  according to their length. 
The  frequencies of cluster occurrence of Fig.~\ref{fig:kullbackshort} (a),(b),(c) are plotted as a function of the cluster duration $\tau_j$  respectively in Figs.~\ref{fig:kullbackshort}(d),(e),(f).  These plots are intended as simplistic illustrations of the probability distribution functions $Q(\tau_j,n)$   and $P(\tau_j,n)$  appearing in the  Eqs.~(\ref{Kullbackdtau},\ref{Kullbackdtaun}). 
\par
The functionals ${ \cal D}(\tau_j,n)$  defined by Eqs.~(\ref{Kullbackdtau}) are estimated for the \textit{realized volatility} series of the financial market data described in Table \ref{tab:data}.
Results for  the S\&P500 market are plotted in Fig.~\ref{fig:kullbackentropy} (a), (b) and (c). The volatility window is $T=180s$.  The volatility is estimated over different monthly horizons ${\cal M}$. Time horizons are ${\cal M}=1 $, ${\cal M}=6$  and ${\cal M}=12$  months respectively  for Figs.~\ref{fig:kullbackentropy} (a)(b)(c). Analogous results have been obtained for the \textit{realized volatility} of the NASDAQ, DJIA, DAX and  FTSEMIB  assets.
The behaviour of the ${ \cal D}(\tau_j,n)$ curves in Fig.~\ref{fig:kullbackentropy} (a)(b)(c) is consistent with the properties expected for relative entropy functionals. In particular,   ${ \cal D}(\tau_j,n)$ is maximum  at small $\tau_j$, then decreases towards the minimum values at large cluster durations ${ \cal D}(\tau_j,n) \rightarrow 0$  as $\tau_j$ increases.
The maximum values of ${ \cal D}(\tau_j,n)$ at small $\tau_j$ is consistent with the loss of power-law correlation and the onset of exponential correlation at cluster duration $\tau_j$ larger than $n$. The divergence components  ${ \cal D}(\tau_j,n)$  plotted in Fig.~\ref{fig:kullbackentropy} take negative values at the small $\tau_j$ pointing to a Hurst exponent  $H_1$ larger than $H_2$. 
\par
 As a further remark, we note that the functionals ${ \cal D}(\tau_j,n)$   are estimated   on the \textit{realized volatility} series of the financial market data, made stationary by the detrending procedure.  The cluster relative entropy described in Section IIA is conditioned by the model probability $Q$,  which  is assumed to be a power law rather than on an unrealistic hypothesis of Gaussianity of returns.
\begin{figure*}%[!h]
\begin{subfigure}[]{0.329 \textwidth}
\includegraphics[width=\textwidth]{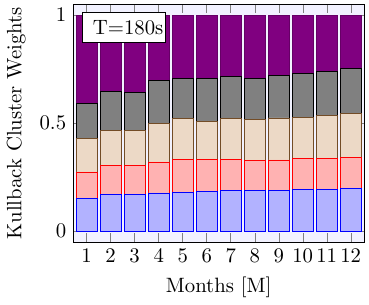}
\caption{}\label{subfig:a}
\end{subfigure}
\begin{subfigure}[]{0.329 \textwidth}
\includegraphics[width=\textwidth]{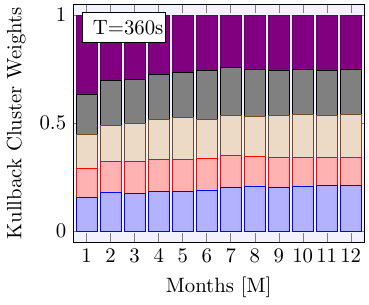}
\caption{}\label{subfig:b}
\end{subfigure}
\begin{subfigure}[]{0.329 \textwidth}
\includegraphics[width=\textwidth]{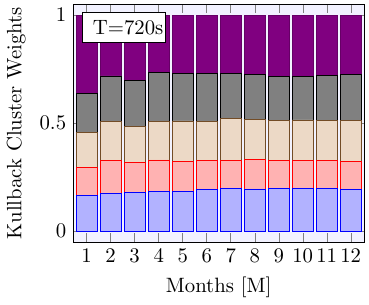}
\caption{}\label{subfig:c}
\end{subfigure}
\caption{\label{fig:kullbackbars} \textbf{Kullback-Leibler weights ${w}_{i,\mathcal{D}}$ vs. investment horizon ${\cal{M}}$ (monthly periods)}.  
 The weights have been estimated by using 
 Eqs.~(\ref{eq:Dindex},\ref{eq:Dweights}) on the realized volatility respectively for the tick-by-tick  series of the markets: S\&P500   (\protect\markerwisteria), NASDAQ (\protect\markerpurple), DJIA (\protect\markerbeige), DAX (\protect\markergray) and FTSEMIB (\protect\markerviolet). The time series of the realized volatility are estimated according to Eq.~(\ref{eq:volatility}) respectively with window $T=180s$ {(a)}, $T=360s$ {(b)} and $T=720s$ {(c)}.}
 \end{figure*}
\par
Next, the weights  ${w}_{i,\mathcal{D}}$ as defined by Eq.~(\ref{eq:Dweights}) are estimated  for the realized volatility   of the five assets.  Results are plotted in  Figs.~\ref{fig:kullbackbars}(a),(b),(c). The weights are estimated for temporal horizons $\cal M$ ranging between one and twelve months. Plots correspond to three values of the volatility window $T=180s$ (a), $T=360s$ (b) and $T=720s$ (c).  At short investment horizons $\cal{M}$ and small volatility windows $T$, the weights deviate from the ${u}_{i}$ values corresponding to uniformly allocated wealth.  As  $\cal{M}$ and $T$ increase, the weights  tend towards the uniform  distribution consistently with the increased level of  surprise with the spreading of cluster duration $\tau_j$. 
A distribution of weights close to ${u}_{i}$, is related to the reduced predictability  at longer horizons and larger volatility windows. This feature can be quantitatively interpreted  bearing in mind that the empirical probability distribution and the model distribution become closer to each other at large cluster durations.   
\par
To  further clarifying the meaning of the results reported  in Fig.~\ref{fig:kullbackentropy} and Fig.~\ref{fig:kullbackbars}, 
the \textit{Shannon cluster entropy} approach has been implemented on the realized volatility of the five markets. 
%%%%%%%%%%%%%
The \textit{Shannon cluster entropy} $\mathcal{S_{C}}[P]$ is obtained by introducing the cluster probability distribution of the \textit{realized volatility} $P(\tau_j,n)$  in the Shannon functional:
\begin{equation}
\mathcal{S}_{j,n}[P]  = -  P(\tau_j, n)\log P(\tau_j,n) \hspace{5pt}.
\label{eq:Shannon}
\end{equation}
Upon summing $\mathcal{S}_{j,n}[P]$ over all the instances  of  cluster duration $\tau_j$ for all the partitions generated by varying the parameter $n$, the \textit{cluster entropy} writes:
\begin{equation}
\mathcal{S_{C}}[P] = -  \sum_{n=1}^N \sum_{j=1}^{{\mathcal N_C}(n)} P(\tau_j, n)\log P(\tau_j,n) \hspace{5pt}.
\label{eq:ShannonC}
\end{equation}
%%%%%%% 
\par 
A  \textit{cluster entropy index} $I_{\cal {S}}$ has been defined by integrating the Shannon cluster entropy functional \cite{ponta2018information,murialdo2021inferring}:
\begin{equation}
I_{\cal {S}}= \sum_{n=1}^N  \sum_{j=1}^m {\cal S} (\tau_j, n) + \sum_{n=1}^N \sum_{j=m}^N {\cal S} (\tau_j, n) \quad,
\label{eq:Sindex}
\end{equation}
Eq.~(\ref{eq:Sindex}) has been written in  Refs.~\cite{ponta2018information,murialdo2021inferring} with the sums over the moving average  $n$ and  lifetime $\tau_j$  implemented  in two steps. To improve readability of Eq.~(\ref{eq:Sindex}),  the  sums  are gathered together, the index $i$ is dropped, and the suffix $\cal S$  is added to distinguish  $I_{\cal S}$  from the index  $I_{\cal D}$  derived in the previous section.
The index defined in  Eq.~(\ref{eq:Sindex}) must be normalized over the ensemble:
\begin{equation}
\label{eq:Sweights}
{w}_{i,\cal {S}}=
\frac{I_{i,\cal {S}}^{}}
{\sum_{i=1}^{{\cal{A}}} 
I_{i,\cal {S}}^{}} \quad ,
\end{equation}
The normalized index are called \textit{Shannon cluster weights} and satisfy  the conditions $\sum_{i=1}^{{\cal{A}}}{w}_{i,\mathcal{S}}=1$  and  $w_{i,\mathcal{S}} \geq 0$. 
By using Eq.~(\ref{eq:Sindex}),  diversity can be quantified  by maximizing  the cluster entropy of \textit{return} and  \textit{volatility} of  the ${{\cal{A}}}$  assets   at  different temporal horizons $\cal{M}$. 
\par
\begin{figure*}%[!h]
\begin{subfigure}[]{0.329 \textwidth}
\includegraphics[width= \textwidth]{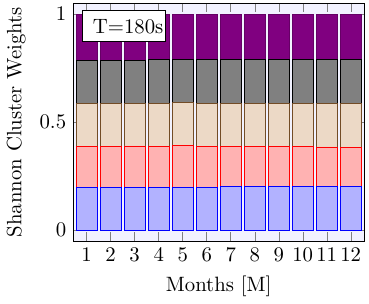}
\caption{}\label{subfig:a}
\end{subfigure}
\begin{subfigure}[]{0.329 \textwidth}
\includegraphics[width= \textwidth]{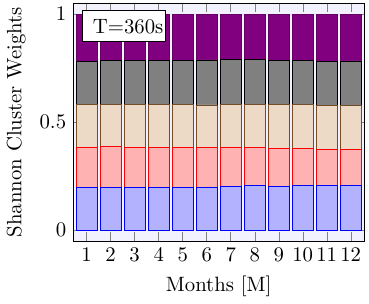}
\caption{}\label{subfig:b}
\end{subfigure}
\begin{subfigure}[]{0.329 \textwidth}
\includegraphics[width= \textwidth]{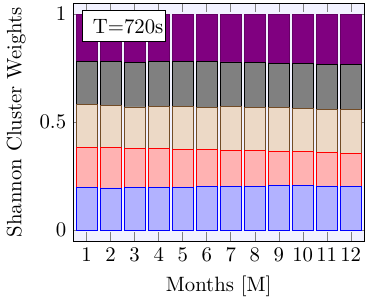}
\caption{}\label{subfig:c}
\end{subfigure}
\caption{\label{fig:shannonbars} \textbf{Shannon cluster weights ${w}_{i,\mathcal{S}}$ vs. investment horizon ${\cal{M}}$ (monthly periods)}.  The weights have been estimated by applying   Eq.~(\ref{eq:Sweights}) to the realized volatility respectively for the tick-by-tick  market  series: S\&P500  (\protect\markerwisteria); NASDAQ (\protect\markerpurple); DJIA (\protect\markerbeige); DAX (\protect\markergray) and  FTSEMIB (\protect\markerviolet).
           The weights refer to the volatility series estimated according to Eq.~(\ref{eq:volatility}) respectively with window $T=180s$ \textbf{(a)}, $T=360s$ \textbf{(b)} and $T=720s$ \textbf{(c)}.   }
 \end{figure*}

The results  obtained by implementing Eq.~(\ref{eq:Shannon}) on the \textit{realized volatility} of the  S\&P500 asset  are shown in Fig.~\ref{fig:shannonentropy}  respectively for the horizons ${\cal M}=1$ (a), ${\cal M} =6$ (b) and ${\cal M} =12$ (c)   months.  The volatility window is $T=180s$ for the three panels. Next, the  weights ${w}_{i,\cal {S}}$ are estimated by using Eq.~(\ref{eq:Sweights}). The weights ${w}_{i,\cal {S}}$  are  shown in  Fig.~\ref{fig:shannonbars} for ${\cal M}=1$  to ${\cal M}=12$ and volatility windows $T=180s$ (a), $T=360s$ (b), $T=720s$ (c). 
%%%%%  MARKOWITZ %%%%%%
\par
\begin{figure*}%[!h]
\begin{subfigure}[]{0.329 \textwidth}
\includegraphics[width=\textwidth]{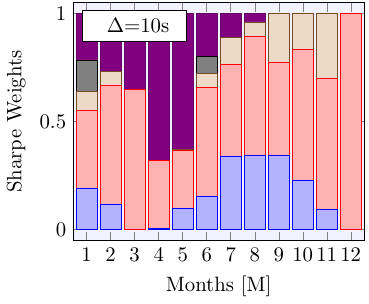}
\caption{}\label{subfig:a}
\end{subfigure}
\begin{subfigure}[]{0.329 \textwidth}
\includegraphics[width=\textwidth]{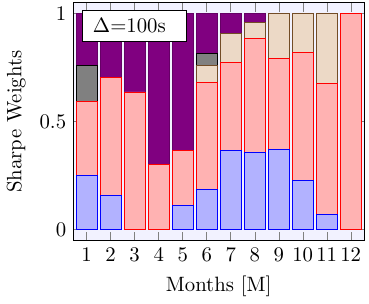}
\caption{}\label{subfig:b}
\end{subfigure}
\begin{subfigure}[]{0.329 \textwidth}
\includegraphics[width=\textwidth]{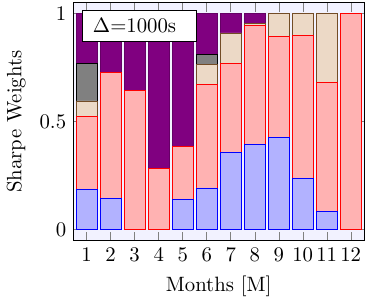}
\caption{}\label{subfig:c}
\end{subfigure}
\caption{\label{fig:sharpebars}  \textbf{Sharpe weights  ${w}_{i}$ vs. investment horizon ${\cal{M}}$ (monthly periods)}.  
 The weights have been estimated  according to the Markowitz approach and Sharpe ratio maximization respectively for the realized volatility of the tick-by-tick data of the markets: S\&P500  (\protect\markerwisteria); NASDAQ (\protect\markerpurple); DJIA (\protect\markerbeige);  DAX (\protect\markergray) and  FTSEMIB (\protect\markerviolet). The three graphs correspond respectively to sampling frequency $\Delta=10s$ {(a)}, $\Delta=100s$ {(b)}, $\Delta=1000s$ {(c)}.}
 \end{figure*}

A comparison with the  weights $w_{i}$  yielded by the traditional mean-variance  Markowitz approach \cite{markowitz52portfolio} is also reported. The weights $w_{i}$ are  chosen to minimize the variance $\sigma^{2}\left(r\right)$  of the return  ${r}$  given the mean $\mu({r})$  by maximizing the \textit{Sharpe ratio}:
\begin{equation}
\label{eq:Sharperatio}
R_{\cal{S}} =\frac{\mu\left(r\right)}{\sqrt{\sigma^{2}\left(r \right)}} \quad .
\end{equation}
The  weights have been maximized on tick-by-tick data of the high-frequency markets described in Table \ref{tab:data} by using the MATLAB Financial Toolboxes \cite{brandimarte2013numerical}. 
Raw market data are sampled to yield equally spaced series with equal lengths. The sampling intervals are indicated by $\Delta$. The Sharpe ratio weights are estimated over  multiple horizons $\cal{M}$ (twelve monthly periods  over the year 2018).
Results  are shown in Fig.~\ref{fig:sharpebars}  respectively for sampling interval $\Delta=10s$ (a), $\Delta=100s$ (b) and  $\Delta=1000s$ (c).  One can note the high variability of the weights of the same asset  over  consecutive periods and  biased distribution  towards  riskiest  assets. Compared to Sharpe ratio weights $w_{i}$, the relative cluster entropy  weights  $w_{i,\mathcal{D}}$ exhibit  less variability, with values smoothly approaching the equally distributed weights $u_{i}$ as increasing investment horizons $ \cal M$  and volatility windows $T$ are considered. 
\par
It is worth remarking that the definition of the weights is not unique and might be modified  to privilege other investment strategies depending upon financial product specifications. 
\par
The weights $w_{i,\mathcal{D}}$ are defined in Eq.~(\ref{eq:Dweights}) in terms of the reciprocal of the entropy index ${I_{i,\cal {D}}^{-1}}$ whereas the weights $w_{i,\mathcal{S}}$ are defined in Eq.~(\ref{eq:Sweights}) in terms of ${I_{i,\cal {S}}}$. This difference is related to the complementary behaviour of the Kullback-Leibler and Shannon cluster entropies, in particular for what concern the dependence on the cluster duration $\tau_j$ at small and large scales. 
\par 
The investment
timeline has been split  into 
%[$\xi$]
periods of equal duration 
$\Theta$ (one month).
%representing the months are considered. Thus, $\xi$ ranges between 1 and 12. If $\xi=12$, $\xi \times \Delta =Y$.
The portfolio is evaluated over multiple time horizons  $ {\cal M}= \xi  \Theta$.  
The different sets of vectors $w_i ({\cal M})$  represent the multi period portfolio which calibrates the long horizon weights over the short ones.
 In the next section, we will consider an investment timeline of one year with $\Theta$ equal to  one month, then $\xi \in {[1,12]}$ yielding investment horizons $ {\cal M}$ lasting from one to twelve months. Investors can choose among different alternatives of $w_i ({\cal M})$   according to their period preferences and the specifity of the investment products.

%%%%%%DISCUSSION %%%%%%%%%% 
\section{Discussion}
\label{Sec:Discussion}
\begin{figure*}%[h]
\begin{subfigure}[]{0.329 \textwidth}
 \includegraphics[width=\textwidth]{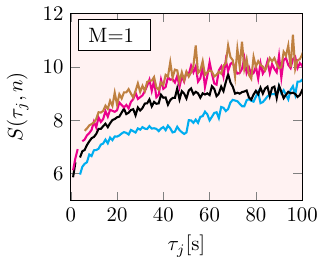}
\caption{}\label{subfig:a}
\end{subfigure}
\begin{subfigure}[]{0.329 \textwidth}
 \includegraphics[width=\textwidth]{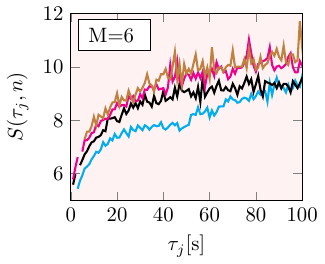}
\caption{}\label{subfig:b}
\end{subfigure}
\begin{subfigure}[]{0.329 \textwidth}
 \includegraphics[width=\textwidth]{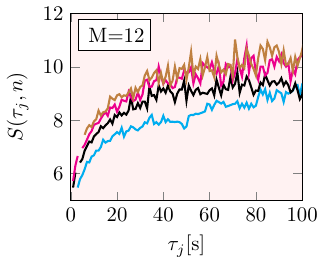}
\caption{}\label{subfig:c}
\end{subfigure}
 \caption{\label{fig:shannonentropy}   \textbf{Shannon cluster entropy functional ${\cal{S}}(\tau_j,n)$   vs cluster duration}. Each curve  is estimated by means of Eq.~(\ref{eq:Shannon})  based on the Shannon functional of  ${P}(\tau_j,n)$ the probability distribution of the cluster duration of the volatility series  of the  S\&P500 asset.
Time series have length $N=492023$. Volatility is estimated according to Eq.~(\ref{eq:volatility}) with window $T = 180s$ for the three graphs. Plots refer respectively to  one (${\cal{M}}=1$){(a)}, six (${\cal{M}}=6$){(b)} and twelve  (${\cal{M}}=12$){(c)}  monthly periods sampled out of the year 2018. Different curves in the same plot refer to different values of  the moving average
window  $n$ ranging from  $50s$ to  $200s$ with step $50s$. 
}
 \end{figure*}

In Section \ref{Sec:Results}, the stochastic volatility process of five assets has been investigated by adopting  the Kullback-Leibler functional $\mathcal{D}_{j,n}[P || Q ] $   and entropy $\mathcal{D}_{C}[P || Q ] $  defined by the Eqs.~(\ref{Kullbackdtau},\ref{Kullbackdtaun}). The \textit{divergence} is quantified between the cluster probability distributions $P$ of the \textit{realized volatility} time series  and  $Q$ of a fully uncorrelated process (a simple Brownian motion with $H=1/2$ taken as reference. Results are shown in Fig.~\ref{fig:kullbackentropy} (a),(b),(c).
For comparison purposes, the realized volatility time series of the five assets have been further analysed by using Eqs.~(\ref{eq:Shannon},\ref{eq:ShannonC}), i.e. the  information theoretic measures based on the Shannon functional $\mathcal{S}_{j,n} [P]$ and entropy $\mathcal{S_{C}}[P]$. Results  for S\&P500 are  shown in Fig.~\ref{fig:shannonentropy}. 
\par
By comparing the curves in  Fig.~\ref{fig:kullbackentropy} with those in Fig.~\ref{fig:shannonentropy}, one can note that the  functionals $\mathcal{D}_{j,n}[P || Q ] $ and $\mathcal{S}_{j,n} [P]$ exhibit a different dependence  on $\tau_j \in [1,\infty]$. More specifically $\mathcal{D}_{j,n}[P || Q ] $  is a decreasing function of  $\tau_j$, while $\mathcal{S}_{j,n}[P]$ increases as $\tau_j$ increases.
The maximum of  $\mathcal{D}_{j,n}[P || Q ] $  occurs at short cluster duration ($\tau_j \rightarrow 1$). Then,  $\mathcal{D}_{j,n}[P || Q ] $ decreases as $P$ approaches the  fully uncorrelated probability distribution at  large cluster duration $\tau_j$.   $\mathcal{D}_{j,n}[P || Q ] $ takes larger values as the probability distribution  $P$ is strongly correlated (power-law correlated) compared to the uncorrelated distribution $Q$ taken as reference.   
Conversely, the Shannon entropy functional  is minimum ($\mathcal{S}_{j,n} [P]\sim \ln 1 = 0$) corresponding to the minimum uncertainty (minimum surprise) on the outcome of the cluster duration $\tau_j$. If $P$ is a  fully developed power-law distribution defined over a large number of cluster sizes, the entropy approaches the maximum value corresponding to maximum uncertainty (maximum surprise) on the outcome at large cluster sizes. The increase is related to  the power-law distribution spreading over a broad range of cluster values being the entropy a measure of the uncertainty (surprise) of the cluster size outcomes. 
In summary, while the Shannon cluster entropy takes its minimum value at short cluster duration $\tau_j$, the Kullback-Leibler divergence is maximum at short lifetime and becomes negligible as cluster duration $\tau_j$ increases. Hence, the two measures  provide complementary information when implemented on the same datasets.
The different dependence of $\mathcal{D}_{j,n}[P || Q ] $ and $\mathcal{S}_{j,n} [P]$  on  $\tau_j$ are reflected in the  Kullback-Leibler $\mathcal{D}_{C}[P || Q ] $ and Shannon $\mathcal{S_{C}}[P]$  entropy, obtained by summing the functionals over the entire ensemble of clusters.

\par
To avoid the impression that the information theoretic measure based on the Eqs.~(\ref{Kullbackdtau},\ref{Kullbackdtaun}) might have only a speculative interest aimed at scrutinizing the roughness of the \textit{realized volatility} time series, a practical metric has been developed leading to definition of the relationships Eqs.~(\ref{eq:Dweights},\ref{eq:Dindex})  for quantifying market risk and related wealth allocation.
The index $I_{\mathcal{D}}$ and the set of weights ${w}_{i,\mathcal{D}}$  are estimated for the high-frequency market data series (details in Table \ref{tab:data}) over multiple consecutive horizons $\cal{M}$. The results are shown in Fig.~\ref{fig:kullbackbars}. 
A continuous set of values of the weights ${w}_{i,\mathcal{D}}$ with a smooth and sound dependence on the horizon $\cal{M}$ can be observed.   
The regularity and smoothness of the entropy-based estimate of the weights  can be related to the stationary detrended distribution rather than the unrealistic mean-variance hypothesis of Gaussianity and stationarity  of the return  of the financial series.  The proposed approach  uses a stationary set of variables (i.e. the detrended clusters and their durations $\tau_j$ of the return and volatility series) rather than non-stationary and not-normal variables as asset returns and volatility. The relevance of the stationarity condition when relative entropy measures are adopted has been discussed in \cite{roldan2012entropy}.
The relative cluster entropy weights ${w}_{i,\mathcal{D}}$  vary continuously and take values close to the equally distributed  weights ${u}_{i}$.
At short investment horizons $\cal{M}$,  the weights ${w}_{i,\mathcal{D}}$  deviate more from the equally distributed values than they do at long investment horizons. 
\par
This behavior is consistent with the structure of the relative cluster entropy functional ${\cal D}(\tau_j,n)$ which maximizes the deviation of the cluster distribution at short $\tau_j$ duration   as opposed to the ${ \cal S}(\tau_j,n)$ functional  which yields the maximum deviation at large $\tau_j$ values. 
Consistently with the different behavior of the functionals  ${ \cal D}(\tau_j,n)$  and ${ \cal S}(\tau_j,n)$, one can note a different dependence of the weights on the volatility window $T$.   The volatility plays a minor role as  the volatility window $T$ increases, i.e. when the functional  ${ \cal D}(\tau_j,n)$ is less sensitive to the distance between the probability distributions $P$ and $Q$. Hence the relative cluster weights ${w}_{i,\mathcal{D}}$ take values closer to the uniform  ${u}_{i}$ distribution with larger volatility windows $T$.
\par
\begin{table*}%[!h]
\footnotesize
\centering
\begin{tabular}{lccccc}
  %  &
      \multicolumn{6}{c}{\textbf{Cluster entropy weights ${w}_{i,\mathcal{D}}$}}
 \\
    \hline
{${\cal{M}}$} & { S\&P500} & {NASDAQ} & {DJIA}  & {DAX}  & {FTSEMIB}  \\
\hline
\hline
1  &    0.2229	 &    0.2707	    &        0.2177      &   	0.2066	        &       0.082	                \\
2  &	0.2075	    &    0.2708	    &        0.2227	     &       0.198	      &           0.101	              \\
3  &	0.2071	    &    0.2659	    &        0.223	      &       0.2044	     &           0.0996	               \\
4  &	0.212	     &0.2627	        &    0.2136	         &       0.1871	     &           0.1247	                 \\
5  &	0.2103	    &    0.253	     &        0.2022	     &       0.2049      &       	0.1296	                      \\
6  &	0.2067	    &    0.2577     &    	0.2153	        &       0.1894      &           	0.1308	                    \\
7  &	0.2009	    &    0.2654	    &        0.2037	     &           0.1945  &           	0.1355	                      \\
8  &	0.1996	    &    0.2742	    &        0.2006	     &           0.1958  &           	0.1298	                        \\
9  &	0.1996	    &        0.2737 &        	0.1959     &       	0.1944	    &               0.1363	                       \\
10 &	0.1985	    &    0.2697	    &            0.2017	 &           0.1887  &               	0.1414	                        \\
11 &	0.1966	    &    0.2759	    &        0.1894	     &           0.1899	 &                   0.1482	                       \\
12 &	0.1934	    &0.2758	        &        0.1894      &           	0.1837	&               0.1577\\
\end{tabular}
\qquad
\begin{tabular}{lccccc}
  %  &
      \multicolumn{6}{c}{\textbf{  Sharpe Ratio  weights ${w}_{i}$}} 
 \\
    \hline
{${\cal{M}}$} & { S\&P500} & {NASDAQ} & {DJIA}  & {DAX}  & {FTSEMIB}  \\
\hline
\hline
1  &    0.1893    &    	0.3611	&         0.0877	     &     0.1475	      &        0.2144	 \\
2  &	0.1187    & 	    0.5508	  &       0.0622	       &     0	             &        0.2683	\\
3  &	0	        &       0.6477	   &         0	            &     0	             &        0.3523	   \\
4  &	0.0061    &       0.3151	  &                 0	    &         0	         &            0.6787	    \\
5  &	0.0994    &       0.2657	 &         0.0062	     &         0	         &        0.6288	\\
6  &	0.1518    &       0.5080	 &         0.0648	      &     0.0770	      &        0.1985	        \\
7  &	0.3392    &       0.4240	 &         0.1261	     &         0	         &        0.1107	          \\
8  &	0.3418    &       0.5513	 &          0.0683	    &     0	             &        0.0386	     \\
9  &	0.3419    &       0.4343	 &         0.2238	     &             0	     &            0	       \\
10 &	0.2269    &       0.6085	 &         0.1645	     &             0	     &            0	          \\
11 &	0.0950    &       0.6070	 &         0.2980	     &             0	     &            0	       \\
12 &	0	        &        1	         &         0	            &         0          &        	0      \\
\end{tabular}
\caption{(Left)  \textbf{Kullback-Leibler cluster entropy weights ${w}_{i,\mathcal{D}}$}; The first column includes the investments horizon ${\cal{M}}$  ranging from one to twelve months.   
 The weights are obtained  by using 
 Eqs.~(\ref{eq:Dindex},\ref{eq:Dweights}) for the realized volatility series  respectively for the tick-by-tick  series of the markets: S\&P500, NASDAQ, DJIA, FTSEMIB and DAX. The  realized volatility is estimated according to Eq.~(\ref{eq:volatility})  with window $T=180s$. (Right)
\textbf{Sharpe Ratio weights (Markowitz) ${w}_{i}$}.  The first column includes the investments horizon ${\cal{M}}$  ranging from one to twelve months.   
The weights have been estimated  according to the Markowitz approach and Sharpe ratio maximization  for the realized volatility of the tick-by-tick data (sampled with frequency  $\Delta=10s$) 
\label{tab:kullbackWeigth}}
\end{table*}
For the sake of clarity, a numerical example of wealth allocation is provided. Let's consider a total allocated wealth $W$ as a net investment after taxation and commissions. The portfolio built using the  Kullback-Leibler weights  ${w}_{i,\mathcal{D}}$ is compared with those obtained by maximizing the Sharpe ratio  ${w}_{i}$  according to Markowitz theory and the uniformly distributed weights ${u}_{i}$.  Two types of investor strategies are considered: 
\begin{description}
\item[a]   \textit{lazy investor} keeping the portfolio shares unchanged over the year. The wealth is allocated the first day of  the year to the weights of the month one ${\cal{M}}=1$. Then the portfolio weights are left unchanged. The portfolio value changes only due to the price changes of the assets.
 \item[b] \textit{active investor} regularly  updating the portfolio shares (monthly in our example). Specifically, the wealth is re-allocated to each index shares according to the updated portfolio weights of the corresponding period ${\cal{M}}$;
\end{description}
In the case of the lazy investor the number of shares of each index in the portfolio does not change  during the year, whereas in the case of the active investor the number of shares of each index is updated every month evaluating the initial wealth allocated to each index (dividing $W_o$ by the weight of the corresponding month) dividing by the price of each index of the corresponding month.
\par
The expected profits are estimated by considering  the adjusted close prices  of the five indexes on the first available  day of each month of the year 2018. The adjusted prices $p_1$, $p_2 \ldots p_{12}$ are given in US dollars and are reported in  Table \ref{tab:data}. The adjusted close prices are publicly available on platforms like Yahho finance, Google finance, Bloomberg.
\par
By using  the uniform weights ${u}_{i}=0.2$ for all $i$ and for all ${\cal{M}}$, the  Kullback-Leibler weights  ${w}_{i,\mathcal{D}}$ (Table \ref{tab:kullbackWeigth} left)  and the Sharpe ratio weights ${w}_{i}$ (Table \ref{tab:kullbackWeigth} right) the initial wealth $W_0=500000 \,\, \text{USD}$ has been allocated at the first month. The number of shares of each index is evaluated by taking the ratio of the amount of wealth allocated to each index to the  price  $p_1$,  at the beginning of the period ${\cal M}=1$.
The value of the portfolio is evaluated each month  considering the number of shares of each index times the price at the end of the month approximated with the price at the beginning of the next month. Table \ref{tab:portfolios} shows the portfolios' evaluated for the six cases, i.e. the two strategies (lazy and active) and three different weights (uniform, Kullback-Leibler  or Sharpe ratio). 
The portfolio value is evaluated as the sum of the wealth allocated to each index in the portfolio.
 The monthly profit is evaluated as the difference between the portfolio value  $W_{\cal M}$ at each month and the initially allocated wealth $W_0$.
Table \ref{tab:profit} shows the profit or loss in the six cases. It is worth noting that the Kullback-Leibler weigths portfolio shows the best performances both with the active and lazy investor strategy compared to the equally weighted and Markowitz portfolio.

%%%%%%%%%%%%% tabelle portafogli con i 5 titoli %%%%%%%%%%%%%%%%
\begin{table*}%[!h]
\footnotesize
\centering
\begin{tabular}{lccccc|c}
   & \multicolumn{6}{c}{\textbf{Uniform portfolio  (Not updated)} }
 \\
{${\cal{M}}$} &SPX	&	NASDAQ	&	DJIA	&	DAX	&	FTSE	&	Portfolio	\\
\hline
\hline
1	&	104680	&	105408	&	105489	&	104801	&	97550	&	517929	\\
2	&	99327	&	102478	&	99133	&	95198	&	97933	&	494071	\\
3	&	95773	&	98047	&	95247	&	94626	&	97979	&	481675	\\
4	&	98478	&	101766	&	97079&	96722	&	100684	&	494731	\\
5	&	101439	&	107812	&	99239&	95350	&	88156	&	491998\\
6	&	101146&	108003	&	97918	&	91844	&	90873	&	489786	\\
7	&	104360	&	109995&	102053	&	95198	&	87263	&	498872	\\
8	&	107452	&	115475	&	104545	&	90929	&	8401	&	502421\\
9	&	108486	&	114705	&	107360	&	91653	&	79058	&	501265	\\
10	&	101652	&	106096	&	102242	&	83803	&	77615	&	471410	\\
11	&	103507	&	106202	&	104038	&	83498	&	75913	&	473160	\\
12	&	93108	&	95133	&	94047&	77134&	77771	&	437195	\\
\vspace{10pt}
\end{tabular}
%\caption{Portfolio value divided among the indexes in the portfolio. spesi uguali no agg
%\label{tab:PUNoAgg}}
%\end{table*}
\quad
%\begin{table*}%[!h]
%\centering
\begin{tabular}{lccccc|c}
    &
      \multicolumn{6}{c}{\textbf{Uniform portfolio (Monthly updated)} }
 \\
{${\cal{M}}$} &SPX	&	NASDAQ	&	DJIA	&	DAX	&	FTSE	&	Portfolio	\\
\hline
\hline
1	&	104680	&	105408	&	105489&	104801	&	97550	&	517929\\
2	&	94886	&	97220	&	93975	&	90836	&	100393	&	477311	\\
3	&	96422	&	95676	&	96079	&	99399	&	100046	&	487625	\\
4	&	102824	&	103792	&	101923	&	102215	&	102760	&	513516	\\
5	&	103006	&	105940	&	102224	&	98581	&	87557	&	497311	\\
6	&	99710	&	100176	&	98668	&	96322	&	103082	&	497961\\
7	&	103177	&	101844	&	104223	&	103651	&	96027	&	508925	\\
8	&	102963	&	104981	&	102442	&	95516	&	96279	&	502183	\\
9	&	100962&	99333&	102692&	100796	&	94097	&	497881	\\
10	&	93700	&	92494	&	95232	&	91434	&	98174	&	471037\\
11	&	101824	&	100100	&	101756	&	99636	&	97807	&	501124	\\
12	&	89953	&	89577	&	90396	&	92377	&	102447	&	464753	\\
\vspace{10pt}
\end{tabular}
%\caption{Portfolio value divided among the indexes in the portfolio. pesi uguali agg  \label{tab:PUAgg}\end{table*}
%\begin{table*}%[!h]
%\centering

\begin{tabular}{lccccc|c}
   % &
      \multicolumn{7}{c}{\textbf{Kullback-Leibler  portfolio (Not updated)} }
 \\
{${\cal{M}}$} &SPX	&	NASDAQ	&	DJIA	&	DAX	&	FTSE	&	Portfolio	\\
\hline
\hline
1		&    116666	&	142670	&	114825	&	108260	&	39995	&	522417	\\
2		&    110700	&	138704&	107907&	98339&	40152	&	495804	\\
3		&    106739	&	132707	&	103676&	97749	&	40171&	481045	\\
4		&    109754	&	137741	&	105671	&	99914&	41280	&	494362	\\
5		&    113054	&	145924	&	108022	&	98497&	36144	&	501642	\\
6		&    112727	&	146182	&	106583	&	94875	&	37258	&	497627	\\
7		&    116309	&	148879	&	111085	&	98339&	35778&	510392	\\
8		&    119756	&	156296	&	113798	&	93930	&	34447	&	518228	\\
9		&    120908 &	155253	&	116862&	94678	&	32414	&	520116\\
10		&   113292	&	143601	&	111291	&	86568	&	31822	&	486576	\\
11		&   115359&	143745	&	113245	&	86253	&	31124	&	489728	\\
12		&   103769	&	128763	&	102370	&	79679	&	31886	&	446469	\\
\vspace{10pt}
\end{tabular}
%\caption{ Portfolio value divided among the indexes in the portfolio. kullback no agg
%\label{tab:kullNoAgg}}
%\end{table*}
\quad
%\begin{table*}%[!h]
%\centering
\begin{tabular}{lccccc|c}
  % &
      \multicolumn{7}{c}{\textbf{Kullback-Leibler portfolio (Monthly updated)} }
 \\
{${\cal{M}}$} &SPX	&	NASDAQ &	DJIA	&	DAX	&	FTSE	&	Portfolio	\\
\hline
\hline
1	&	116666	&	142670	&	114825	&	108260	&	39995	&	522417	\\
2	&	98444	&	131636	&	104641	&	89928	&	50698	&	475348	\\
3	&	99845	&	127202	&	107128	&	101586&	49823	&	485586	\\
4	&	108993	&	136332	&	108854	&	95622&	64071	&	513873	\\
5	&	108311	&	134015&	103349	&	100996	&	56737	&	503409	\\
6	&	103051	&	129077	&	106216	&	91217	&	67415	&	496979	\\
7	&	103642	&	135147	&	106151	&	100801 &	65058	&	510801	\\
8	&	102757	&	143930	&	102749	&	93510	&	62485	&	505432	\\
9	&	100760	&	135937	&	100587	&	97974&	64127	&	499386	\\
10	&	92998	&	124728	&	96042	&	86268	&	69409	&	469447	\\
11	&	100093	&	138088	&	96362	&	94604	&	72475	&	501624	\\
12	&	86984	&	123527	&	85605	&	84849	&	80779	&	461747	\\
\vspace{10pt}
\end{tabular}
%\caption{ Portfolio value divided among the indexes in the portfolio. kullback agg
%\label{tab:kullAgg}}
%\end{table*}
%\begin{table*}%[!h]
%\centering
\begin{tabular}{lccccc|c}
  %  &
      \multicolumn{7}{c}{\textbf{Sharpe ratio portfolio (Not updated)} }
 \\
{${\cal{M}}$} &SPX	&	NASDAQ	&	DJIA	&	DAX	&	FTSE	&	Portfolio	\\
\hline
\hline
1	&	99057	&	190305	&	46281	&	77311	&	104561	&	517517	\\
2	&	93992	&	185016&	43493&	70226	&	104972	&	497700	\\
3	&	90629	&	177017	&	41788	&	69805	&	105021	&	484261	\\
4	&	93189	&	183731	&	42592	&	71351	&	107920	&	498784	\\
5	&	95991	&	194646	&	43539	&	70339	&	94492	&	499009	\\
6	&	95713	&	194990	&	42960	&	67752	&	97405	&	498822	\\
7	&	98755&	198587	&	44774	&	70226&	93535	&	505880	\\
8	&	101681	&	208481	&	45867&	67078	&	90055	&	513164	\\
9	&	102659	&	207091	&	47102	&	67612&	84740&	509206	\\
10	&	96193	&	191547&	44857&	61820&	83193	&	477613	\\
11	&	97948	&	191739&	45645	&	61595	&	81369	&	478299	\\
12	&	88107	&	171756	&	41261	&	56901	&	83361	&	441388	\\
\end{tabular}
%\caption{Portfolio value divided among the indexes in the portfolio. Sharpe no agg
%\label{tab:sharpeNoAgg}}
%\end{table*}
\qquad
%\begin{table*}%[!h]
%\centering
\begin{tabular}{lccccc|c}
    %&
      \multicolumn{7}{c}{\textbf{Sharpe ratio  portfolio (Monthly updated)} }
 \\
{${\cal{M}}$} &SPX	&	NASDAQ	&	DJIA	&	DAX	&	FTSE	&	Portfolio	\\
\hline
\hline
1	&	99057	&	190305	&	46281&	77311	&	104561&	517517	\\
2	&	56336	&	267728	&	29242	&	0	&	134653	&	487961	\\
3	&	0	&	309845	&	0	&	0	&	176236	&	486081	\\
4	&	3153	&	163536	&	0	&	0	&	348741&	515431	\\
5	&	51189	&	140727	&	3157&	0	&	275263	&	470337	\\
6	&	75656&	254453	&	31943	&	37075&	102314	&	501442	\\
7	&	174985	&	215907	&	65719	&	0	&	53152&	509764	\\
8	&	175966	&	289390	&	34977	&	0	&	18577	&	518912	\\
9	&	172592&	215725	&	114889	&	0	&	0	&	503208	\\
10	&	106321&	281430	&	78342	&	0	&	0	&	466094\\
11	&	48378	&	303812	&	151596	&	0	&	0	&	503787	\\
12	&	0	&	447888	&	0	&	0	&	0	&	447888	\\

\end{tabular}
\caption{Wealth montly allocated to the five indexes. The total wealth allocated is reported in the last column.
\label{tab:portfolios}}
\end{table*}

\begin{table*}%[!h]
\footnotesize
\centering
\begin{tabular}{lcccccc}
    &   \multicolumn{6}{c}{\textbf{Profit} }\\
   &   \multicolumn{2}{c}{{ Uniform} } &      \multicolumn{2}{c}{{Kullback-Leibler} } &      \multicolumn{2}{c}{{Sharpe ratio} }\\
{${\cal{M}}$} & { (a)} & {(b)} & { (a)} & {(b)}  & { (a)} & {(b)}   \\
\hline
\hline
1	&	17929	&	17929	&	22417	&	22417	&	17517	&	17517	\\
2	&	-5928	&	-22688	&	-4195	&	-24651&	-2299	&	-12038	\\
3	&	-18324&	-12374&	-18954	&	-14413	&	-15738	&	-13918	\\
4	&	-5268&	13516&	-5637	&	13873&	-1215	&	15431\\
5	&	-8001&	-2688&	1642&	3409	&	-990	&	-29662\\
6	&	-10213	&	-2038	&	-2372&	-3020	&	-1177&	1442	\\
7	&	-1127	&	8925	&	10392	&	10801	&	5880&	9764\\
8	&	2421	&	2183	&	18228&	5432	&	13164	&	18912	\\
9	&	1265	&	-2118&	20116	&	-613&	9206	&	3208	\\
10	&	-28589	&	-28962	&	-13423&	-30552&	-22386	&	-33905\\
11	&	-26839	&	1124	&	-10271&	1624&	-21700	&	3787\\
12	&	-62804	&	-35246	&	-53530&	-38252	&	-58611&	-52111	\\
\hline
\hline
Year	&	-145480	&	-62438&	-35589	&	-53944	&	-78351	&	-71571\\
\end{tabular}
\caption{ Profit or loss of the six portfolios considered. The values are evaluated as the difference between the portfolios' values and the initial quantity of wealth invested.
\label{tab:profit}}
\end{table*}

\section{Conclusions}
\label{Sec:Conclusions}
In this work, we have shown how the relative cluster entropy can quantify the divergence between the probability distribution  of the coarse grained time series  of the  realized volatility defined by Eq.~(\ref{eq:volatility}) and a reference probability distribution  estimated over a model time series (a simple random walk with $H=0.5$). The volatility series is estimated for each market (S\&P500, DJIA, NASDAQ,  DAX, FTSEMIB) over multiple time scales $\tau_j$ and investment periods $ \cal M$. The results are consistent with a value of the correlation exponent $H>0.5$. There is a great interest in the financial research community in  investigating volatility as a fractional Brownian motion and in particular whether the volatility can be considered a rough/smooth process  see the discussion in Ref. \cite{comte1998long,gatheral2018volatility}.
\par 
A dynamic  investment strategy has been proposed  by exploiting the features  of the relative cluster entropy approach.  Importantly, the proposed approach of portfolio construction does not rely on flawed assumptions  on the market data such as  normal and stationary distributions needed to apply the Markowitz and Sharpe ratio approach. 
\par 
It is worth remarking that the current study differs from the analysis reported in Ref.~\cite{carbone2022relative} mainly addressed to demonstrate the relative cluster entropy methodology.  Then as a case study the approach has been implemented on  price series that are commonly considered to behave as simple uncorrelated Brownian motions with $H\sim 0.5$.  
\par
The outcome of this work  relies on the Kullback-Leibler divergence via coarse graining of sequences of power-law distributions of fractional stochastic processes with specific focus on the time series of the realized volatility.   The Kullback-Leibler divergence is a pillar of the classical and quantum information theory (\cite{vedral2002role,kowalewska2012kullback}). Extensions of this research can be envisioned in different directions.  Other distance measures between probability distributions have been proposed, for example the Malanobis distance, which can be used to analyse the probability distributions of sequence clustering. Another generalization of  interest is the application of the proposed information  measure to chaotic dynamics \cite{asuke2023analysis}. One estimator of chaotic system stability  is the Allan variance $\sigma_{Allan}$ (e.g. in fault detection, epileptic seizures 
from electroencephalography (EEG), stability of electronic oscillators, etc)  a coarse grained multiscaled mapping of the sequences. An interesting future
topic might be the application of the relative cluster entropy and  minimum relative entropy principle to the probability distribution function $P(\sigma_{Allan})$.

\acknowledgments{
This work received financial support from the TED4LAT project (a WIDERA initiative within the Horizon Europe Programme, Grant Agreement: 101079206).}

%\bibliographystyle{unsrt}
%\bibliography{Information2020,portfolios,biblioXreferee}

\end{document}